# Auto-grading for 3D Modeling Assignments in MOOCs

Swapneel Mehta
Dept. of Computer Engineering
D. J. Sanghvi College of Engg.
Mumbai, India
swapneel.mehta@djsce.edu.in

Sameer Sahasrabudhe
Dept. of Computer Science and Engineering
Indian Institute of Technology Bombay, Powai
Mumbai, India
samss@it.iitb.ac.in

Chirag Raman
Language Technologies Institute
Carnegie Mellon University
Pittsburgh, USA
chirag.raman@cs.cmu.edu

Nitin Ayer
Dept. of Computer Science and Engineering
Indian Institute of Technology Bombay, Powai
Mumbai, India
ayernitin@gmail.com

*Abstract*—Bottlenecks such as the latency in correcting assignments and providing a grade for Massive Open Online Courses (MOOCs) could impact the levels of interest among learners. In this proposal for an auto-grading system, we present a method to simplify grading for an online course that focuses on 3D Modeling, thus addressing a critical component of the MOOC ecosystem. Our approach involves a live auto-grader that is capable of attaching descriptive labels to assignments which will be deployed for evaluating submissions. This paper presents a brief overview of this auto-grading system and the reasoning behind its inception. Preliminary internal tests show that our system presents results comparable to human graders.

## I. Introduction

MOOCs have seen considerable interest and have come from being a passive learning mode to one of the primary platforms for the dissemination of knowledge pertaining to cutting-edge technology. Right from the year 2012, this sector has seen a rapid boom, with case studies ranging from Prof. Andrew Ng's platform, Coursera, and Prof. Sebastien Thrun's venture, Udacity [3]. For the purpose of this paper, we will focus on the Open edX platform, specifically IITBombayX and edX, which host the iterations of the 3D Animation and 3D Visualization courses offered to thousands of learners cumulatively, over the period of a few years. Our observations as staff and instructor(s) for these courses have resulted in the motivation for this research and development of such a tool in an effort to improve and enhance the experience of a learner with our course.

## II. The Course

IITBombayX has offered a variety of courses on different domains. While it covers a broad base in order to allow students to make the most of this digital channel, it concurrently provides a series of courses aimed at addressing shortcomings in the pedagogy adopted by instructors across the country. Further, the concept of Blended MOOCs was tested out [4, 5] in an attempt to bring about a reduction in massive attrition rates among learners, and provide an increased sense of collaboration in an otherwise virtual environment. While our course(s) on IITBombayX follow similar pedagogy, the auto-grading of assignments is another approach we propose to further address the factors that seem to impact learner interaction with the offered course. The course to be utilised for the purpose of this test is a 3D Visualisation course to be offered on the edX platform, with approximately 500-700 learners that have signed up for the offering as of two weeks prior to the release.

## III. Motivation

It is intuitive to acknowledge that the average learner relies greatly upon individual motivation in successfully completing a MOOC [1]. As an instructor, then, it becomes a responsibility to engage the students in an environment that is both challenging and enriching. In the light of the analytical data available across most platforms today, the onus is on the course staff to adopt the best practices moving forward [2]. The question of assessments plays a critical role in this setup, and while peer-grading has been explored, it is not difficult to fathom why it poses serious problems when expected to scale [8]. We propose a tool that addresses our problem in a manner that can not only scale but also capture data from submitted assignments that can then be used to improve the nature of problems in an effort to address common areas of weakness on the part of the learners. While initially deployed to follow a single set of rubrics for grading assignments limited to objective parameters over subjective knowledge, it will be built upon to incorporate a multi-stage pipeline for the evaluation of assignments of a more complex, multi-faceted nature. The assignment to be graded in this case, is that of a crown, as demonstrated in Figure 1. The crown is a result of the extrusion of alternate surfaces of the Torus, one of the primitive types of shapes available in Blender, an open-source 3D Modeling software [9]. Due to the reduced complexity of this assignment in comparison with other models expected of the learner, we propose to integrate an auto-grading system that would greatly reduce the manual effort required to grade such submissions individually.

## IV. PROBLEM DEFINITION

Automatic grading of 3D Modeling Assignments has been the focus of much research which has brought about development of tools in the fields of computer science, The conventional problem(s) associated with attaching a measurable label to a 3D Modeling assignment has been associated with the arrival at a formal metric to assess aesthetic value. While there has been research in this area, and a formal weighted metric defined by some universities that offer graduate courses in this domain such as the First School of Architecture of the Politecnico di Torino, Italy [6, 7], this rule-based metric is difficult to implement in a general context especially in cases such as ours where creativity and imagination form a crucial step within the learning path. In these papers the authors present a specific subset of parameters that have to be adhered to in order for a submission to be graded. There are points allotted for each parameter and a failure to meet the expected level of proficiency results in a deduction from the maximum score. This serves as a useful paradigm in assessing proficiency in 3D Modeling. However when a course encourages visualization, creativity, and novelty, it becomes exponentially difficult to arrive at a subset of such metrics, or even to expect thousands of precocious learners to adhere to such a set of rules. We have empirically found that a formal set of rules such as the fixed position of an object or camera in a submission is difficult to expect and ultimately evaluate when learners rank among the thousands.

Secondly, the aesthetic factor gains more weight in the context of our course on 3D Visualization which encourages aesthetic freedom, including customisation of materials, texture, color-scheme, and as a result, receives a wide array of novel submissions that range from the expected to the amazing. Providing the feedback for such assignments currently involves a human-in-the-loop procedure, with the grades often serving as an informal portion of the course. However, the observed phenomenon has been that in spite of the optional nature of some assignments, they are duly submitted, and feedback welcomed by learners both via email as well as on the discussion forum for this course. In such a scenario, we feel that limiting the scope for submissions, by introducing a rule-based submission procedure will negatively impact the learner's enthusiasm for the course.

## V. AUTOGRADING TOOL

The autograding tool assesses the submissions by comparing them with an 'ideal' submission called a 'rubric'.

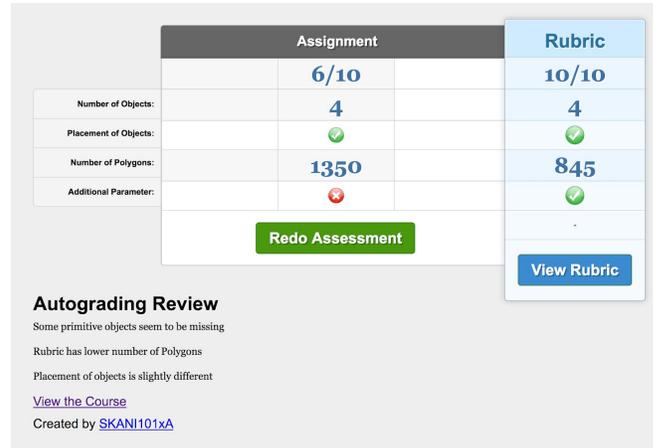

Figure 1. Sample Autograding by the Assessment Tool

### A. Some Common Mistakes

The primitive object type used is incorrect; a crown is often seen made from a cube or sphere.
There is unnecessary complexity introduced into the submission by adding surface modifiers; extrusions and smoothening of surfaces.
Incorrectly extruded planes; the process outlined is the extrusion of alternate plane surfaces while the submissions do not heed this and extrude random surfaces.
Camera is incorrectly placed, leading to an incomplete render of the actual model.
Submissions are often incomplete, or copied from other participants.

### B. Assessment Parameters

- We utilise the location and rotation of an object in order to determine the similarity to the original pose expected for the object to be in. Since there is a possibility for the object to be in a rotated scale, we allocate a lower weight to this parameter.
- Another parameter we consider is the scale of the object in the submission. If the scale varies by a large factor, a negative mark is allocated and the overall grade reduces. In cases where scale is subjective, a lower weight to this parameter would result in a more 'human-like' grade.
- Finally, we check the number of polygons in the NIsubmission and verify it's ratio to the number of polygons in the rubric. Permitting an error-band, we subtract a grade if there is a wide disparity in this ratio i.e. if it lies beyond the [0.7, 1.3] range.

- There is a check for the number of objects and type of objects that have been utilized in the submission, and the wrong primitive object type would carry a higher weight, contributing to a greater negative mark.

*C. Sample Submissions*

- Figure 2 depicts the incorrect usage of a primitive (left) and the improper placement of a camera (right) within a submission.
- Figure 3 presents a correct, but incomplete submission of the crown - only a few of the faces have been extruded; the rendered image of the rubric has been provided for comparison.

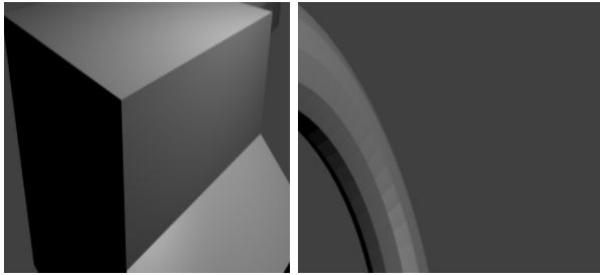

Figure 2. A wrongly extruded crown built using a cube; an improperly placed camera (right).

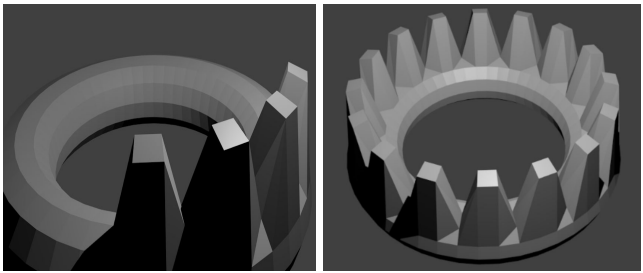

Figure 3. A correct, but incomplete submission; and the rubric (right).

## VI. Deployment

The autograding tool has been deployed as an external standalone tool capable of grading assignments relative to a rubric. This offloads the analysis and reduces the load for the Open edX system. By doing this, we also get the opportunity to maintain a record of submission data that can be used to address the pain points within the provided submissions. Further, we provide a natural language comparison for learners to intuitively understand improvements to their model as shown in Figures 5 and 6.

## VII. Future Work

Recent literature, and investigations into learner engagements within MOOCs has established a strong link between the modes of assessment employed and the attrition rates [11, 12]. Specifically, the provision of actionable feedback boosts learner interaction and eventually completion rates. While auto-grading has a positive correlation with most MOOCs, we highlight the descriptive feedback that it offers to the user in the context of ensuring a positive experience.

There are understandably a number of subjective factors to consider such as the perceived 'difficulty' of a course, and the importance attached to successful submission of assignments for the MOOC that also impact learner engagement. We consider 3D Visualization as falling into the 'easy-to-medium' category and assignments are not enforced within the course, although quizzes are. Thus, the introduction of an auto-grading pipeline within this course would be an efficient approach towards attempting to boost learner engagement through the use of open technology.

References

[1] Barba PD, Kennedy GE, Ainley MD. The role of students' motivation and participation in predicting performance in a MOOC. Journal of Computer Assisted Learning. 2016 Jun 1;32(3):218-31.

[2] Reich, J., 2015. Rebooting MOOC research. *Science*, *347*(6217), pp.34-35.

[3] Pappano, L., 2012. The Year of the MOOC. *The New York Times*, *2*(12), p.2012.

[4] Yousef, A.M.F., Chatti, M.A., Schroeder, U. and Wosnitza, M., 2015. A usability evaluation of a blended MOOC environment: An experimental case study. *The International Review of Research in Open and Distributed Learning*, *16*(2).

[5] Bruff, D.O., Fisher, D.H., McEwen, K.E. and Smith, B.E., 2013. Wrapping a MOOC: Student perceptions of an experiment in blended learning. *Journal of Online Learning and Teaching*, *9*(2), p.187.

[6] Lamberti, F., Sanna, A., Paravati, G. and Carlevaris, G., 2014. Automatic grading of 3d computer animation laboratory assignments. *IEEE Transactions on Learning Technologies*, *7*(3), pp.280-290.

[7] Sanna, A., Lamberti, F., Paravati, G. and Demartini, C., 2012. Automatic assessment of 3D modeling exams. *IEEE Transactions on Learning Technologies*, *5*(1), pp.2-10.

[8] Shah, N.B., Bradley, J., Balakrishnan, S., Parekh, A., Ramchandran, K. and Wainwright, M.J., 2014. Some scaling laws for MOOC assessments. In *KDD Workshop on Data Mining for Educational Assessment and Feedback (ASSESS 2014)*.

[9] The Blender Project, https://www.blender.org/

[10] Severance, C., Hanss, T. and Hardin, J., 2010. Ims learning tools interoperability: Enabling a mash-up approach to teaching and learning tools. *Technology, Instruction, Cognition and Learning*, *7*(3-4), pp.245-262.

[11] Jordan, K., 2015. Massive open online course completion rates revisited: Assessment, length and attrition. *The International Review of Research in Open and Distributed Learning*, *16*(3).

[12] Rai, L., Yue, Z., Yang, T., Shadiev, R. and Sun, N., 2017, August. General impact of MOOC assessment methods on learner engagement and performance. In *Ubi-media Computing and Workshops (Ubi-Media), 2017 10th International Conference on* (pp. 1-4). IEEE.